THE LATEX FILE FOLLOWS:

\documentstyle[art12,bezier]{article}

\begin{document}

\renewcommand{\thefootnote}{\fnsymbol{footnote}}

\thispagestyle{empty}

\hfill \parbox{45mm}{
{\sc CTP\ \#2225} \par
July 1993}

\vspace*{15mm}

\begin{center}
{\LARGE Wilson Loops \par
        in Four-Dimensional Quantum Gravity.}

\vspace{22mm}

{\large Giovanni Modanese}%
\footnote{On leave from University of Pisa, Pisa, Italy. This work is
supported in part by funds provided by the U.S. Department of Energy
(D.O.E.) under contract \#DE-AC02-76ER03069.}

\medskip

{\em Center for Theoretical Physics \par
Laboratory for Nuclear Science \par
Department of Physics \par
Massachusetts Institute of Technology \par
Cambridge, Massachusetts, 02139, U.S.A.}

\bigskip

\medskip

\end{center}

\vspace*{20mm}

\renewcommand{\thefootnote}{\arabic{footnote}}
\setcounter{footnote} 0
\begin{abstract}
A Wilson loop is defined, in 4-D pure Einstein gravity, as the trace of the
holonomy of the Christoffel connection or of the spin connection, and its
invariance under the symmetry transformations of the action is showed
(diffeomorphisms and local Lorentz transformations). We then compute the loop
perturbatively, both on a flat background and in the presence of an external
source; we also allow some modifications in the form of the action, and test
the action of ``stabilized'' gravity. A geometrical analysis of the results in
terms of the gauge group of the euclidean theory, $SO(4)$, leads us to the
conclusion that the correspondent statistical system does not develope any
configuration with localized curvature at low temperature. This ``non-local''
behavior of the quantized gravitational field strongly contrasts with that of
usual gauge fields. Our results also provide an explanation for the absence of
any invariant correlation of the curvature in the same approximation.

\bigskip \bigskip

\end{abstract}

\newcommand{\beq}{\begin{equation}}    \newcommand{\la}{\langle}
\newcommand{\eeq}{\end{equation}}      \newcommand{\ra}{\rangle}
\newcommand{\beqa}{\begin{eqnarray}}   \newcommand{\pa}{\partial}
\newcommand{\eeqa}{\end{eqnarray}}     \newcommand{\half}{\frac{1}{2}}

\newcommand{\m}{\medskip}

\section{Introduction.}
\label{introd}

One open issue of fundamental interest in (3+1)-dimensional quantum gravity is
the identification of a set of meaningful observable quantities.

Either if we regard quantum gravity as a (not yet established!) fundamental
theory, or as an effective quantum field theory which has General Relativity as
its classical limit and goes to some more fundamental theory at short
distances, the observable quantities are important in guiding the
research.

Just because a complete quantum theory of gravity is still lacking, it is not
possible to define exactly what an observable is. The task is particularly
difficult also due to the huge invariance group of gravity, namely the group of
the diffeomorphisms. We shall agree to consider a quantity as an observable, if
the corresponding classical quantity is a geometrical invariant, i.e., it is
invariant under arbitrary transformations of the coordinates.

The observables we intend to study in the present paper are the Wilson loops,
or ``holonomies'', of the Christoffel connection or of the spin connection
$\Gamma^a_{\mu b}$. We have chosen to work in a formalism based on the metric
(or on the vierbein), because we want to assure that the ``size'' and the
``form'' of a loop can eventually be specified through invariant distances and
angles. This seems us to be an essential requirement for a physically
meaningful holonomy; otherwise it would only have topological content, and all
the remaining information about the manifold would be lost.

Generally speaking, it is known that the quantum averages of the loop operators
have to satisfy the analogues of the Migdal-Polyakov loop equations
\cite{Polyakov}. Some general features of these equations have been studied by
Makeenko and Voronov \cite{Voronov}, considering the Christoffel connection and
the usual Einstein action in the metric formalism. What we shall do is simpler
but more explicit. Keeping the local fields as the fundamental dynamical
variables, we shall compute the loops in a few different cases, in order to
learn about their behaviour and their geometrical meaning. The latter turns out
to be quite different from that of the holonomies of Yang-Mills fields.

Our calculations are based on the Einstein-Hilbert action. We shall see,
however, that certain properties of the loops do not depend on the detailed
form
of the action.

Since we work essentially in perturbation theory, some problems like the lower
unboundedness of the euclidean Einstein action do not strictly affect our
results. Nevertheless, the formalism we develop will also lead us to consider,
in the last Section, a different ``source'' for the dynamics of the euclidean
gravitational field, namely the ``stabilized action'' of Greensite.

\m
The plan of the paper is the following. In \S\ \ref{defini} we define
geometrically in detail the Wilson loop of the Christoffel connection and of
the spin connection (in the vector representation) and show their equivalence.

In \S\ \ref{dynami} classical and quantum dynamics are introduced. We also
recall the well known fact that Einstein's action is locally invariant under
$SO(3,1)$, but not under $ISO(3,1)$; so the invariant Wilson loops are just
those of the Lorentz connection, and not, like in (2+1)-gravity, those of a
generalized connection which contains the generators of the translations.

In \S\ \ref{clacas} we give one illustrative example of a classical holonomy,
computing it along a circle of constant radius in a Schwartzschild metric.

In \S\ \ref{smaqua} we consider the case of a weak gravitational field,
quantized around a flat background. We briefly review the corresponding
perturbation theory and prove that the leading contribution to the Wilson loop,
proportional to $\hbar \kappa^2$, vanishes for quite general dimensional and
symmetry reasons.

In \S\ \ref{nonfla} an expression is given for the contribution of order
$\hbar$
to the holonomies computed on a non-flat background. In general this
contribution is not vanishing in that case, due to the lower symmetry of the
background; however, it is only of order $\hbar \kappa^3$.

In \S\ \ref{geoint} we work out in detail the geometrical meaning of the matrix
${\cal U}$ of the parallel transport in the euclidean theory and conclude that
its trace -- that is, the loop ${\cal W}$ -- is the sum of the squares of two
angles, describing an $SO(4)$ rotation. So the vanishing of $\la {\cal W}
\ra_0$ to order $\hbar$ implies that in the equivalent statistical system there
are no excitations with localized curvature at low temperature. This quite
unexpected physical picture also explains the absence of any invariant
correlation of the curvature in this approximation \cite{correl}.

Finally, in \S\ \ref{stagra} we consider a recently proposed ``stabilized''
version of euclidean quantum gravity \cite{greens} and show that the basic
property of the holonomies found in \S\ \ref{smaqua} is maintained in this
case. \S\ \ref{conclu} contains our conclusions.


\section{Definitions.}
\label{defini}

In the so-called ``second order'' (or metric) formalism, classical spacetime is
described by a Lorentzian manifold $M$, whose geometry is encoded in a metric
tensor $g_{\mu \nu}(x)$ of signature $(-1,\, 1,\, 1,\, 1)$ (our conventions are
those of Weinberg \cite{weinbe}).

There is a natural definition of parallel transport of tensors on $M$. For
instance, the variation of a vector $V^\alpha$ by an infinitesimal displacement
$dx^\mu$ is defined by
\beq
  dV^\alpha = - \Gamma^\alpha_{\mu \beta}(x) \, V^\beta \, dx^\mu ,
\label{fra}
\eeq
where $\Gamma^\alpha_{\mu \beta}$ is the Christoffel connection
\beq
  \Gamma^\alpha_{\mu \beta} = \frac{1}{2} \, g^{\alpha \gamma} \,
  \left(\partial_\mu g_{\beta \gamma} + \partial_\beta g_{\mu \gamma} -
  \partial_\gamma g_{\mu \beta} \right) .
\label{ces}
\eeq
Integrating (\ref{fra}) we find that the parallel transport of $V$ along a
finite differentiable curve connecting the points $x$ and $x'$ is performed by
the matrix
\beq
  {\cal U}^\alpha_{\beta}(x,x') = \mbox{P} \, \exp \int_{x}^{x'}
  dy^\mu \, \Gamma^\alpha_{\mu \beta}(y),
\label{cke}
\eeq
where the symbol P means that the matrices
$(\Gamma_\mu)^\alpha_{\beta}=\Gamma^\alpha_{\mu \beta}$ are ordered along
the path. The indices of ${\cal U}^\alpha_\beta(x,x')$ are lowered and
raised by $g_{\alpha \gamma}(x)$ and $g^{\beta \gamma}(x')$, respectively.

When the manifold is curved, the matrix ${\cal U}$ depends not only on the
end points $x$ and $x'$, but also on the path. In fact,
if $C$ is a smooth closed curve on $M$, we define the loop functional
(or ``holonomy'') ${\cal W}(C)$ as
\beq
  {\cal W}(C) = -4 + {\rm Tr} \, {\cal U}(C) =
  -4 + {\rm Tr} \, {\rm P} \exp \oint_C dx^\mu \Gamma_\mu(x) .
\label{fuy}
\eeq

We make the physical requirement that the curve $C$ should be possibly defined
in an intrinsic way (that is, independently of the coordinates), and that its
form and size should be eventually specified by invariant distances and angles.
We also remind that we are not interested here in self-intersecting loops,
non-trivial topologies or global effects, but only in ``local'' effects. Our
attitude should be different, of course, in analyzing the (2+1)-dimensional
theory, where there are no local degrees of freedom.

The term $-4$ in eq.\ (\ref{fuy}) sets the holonomy to zero in the case of a
flat space, when the matrix ${\cal U}$ reduces to an identity matrix.

\m
Under a coordinates transformation $x \to \zeta$, the matrix \ ${\cal U}$
transforms in the following way
\beq
  {\cal U}^{\alpha}_{\beta}(x,x') \to {\cal U}^\alpha_\beta(x,x') \,
  \left[ \frac{\partial {\zeta}^{\gamma}}{\partial {x}^\alpha} \right]_x \,
  \left[ \frac{\partial {x}^\beta}{\partial {\zeta}^{\varepsilon}}
\right]_{x'}.
\eeq
For a closed curve, this transformation, being of the form
\beq
  {\cal U} \to \Omega {\cal U} \Omega^{-1} ,
\label{lau}
\eeq
does not affect the trace of ${\cal U}$. So the loop ${\cal W}(C)$ is invariant
with respect to coordinate transformations.

\m
Instead of the metric formalism, it is also possible to use a ``first order''
formalism, by introducing the vierbein field $e^a_\mu(x)$ and its inverse
$E^\mu_a(x)$ \cite{hehl}, which satisfy the relations
\beqa
  & e^\mu_a(x) E^a_\nu(x) = \delta^\mu_\nu ; \qquad
  e^\mu_a(x) E^b_\mu(x) = \delta^b_a ; & \label{jhy} \\
  & E^a_\mu(x) E^b_\nu(x) \eta_{ab} = g_{\mu \nu}(x) . \label{lcy} &
\eeqa

Any vector $V^\mu$ (or, more generally, any tensor) can be referred to the
vierbein, with ``anholonomic'' components $V^a$ given, in any point $x$, by
\beq
  V^a = V^\mu \, e^a_\mu(x) .
\eeq

The equivalent of (\ref{fra}) in terms of the anholonomic connection
$\Gamma^a_{\mu b}$ is
\beq
  dV^a = - \Gamma^a_{\mu b}(x) V^b dx^\mu
\label{xnc}
\eeq
and the matrix ${\cal U}$ of the finite parallel transport has an expression
which is formally the analogue of (\ref{cke}), namely
\beq
  {\cal U}^a_b(x,x') = \mbox{P} \, \exp \int_{x}^{x'}
  dy^\mu \, \Gamma^a_{\mu b}(y) .
\eeq

Using (\ref{fra}), (\ref{xnc}) and (\ref{jhy}) it is straightforward to verify
that the relation between the connections $\Gamma^\alpha_{\mu \beta}$ and
$\Gamma^a_{\mu b}$ is the following
\beq
  \Gamma^\alpha_{\mu \beta} = E^\alpha_a e^b_\mu \Gamma^a_{\beta b}
  + E^\alpha_a \pa_\beta e^a_\mu
\eeq
and that the relation between the matrices ${\cal U}^\alpha_\beta$
and ${\cal U}^a_b$ is
\beq
  {\cal U}^a_b(x,x') = e^a_\alpha(x) {\cal U}^\alpha_\beta(x,x') E^\beta_b(x').
\label{kxq}
\eeq

It is known that gravity in the vierbein formalism has a local Lorentz
invariance, since eq.\ (\ref{lcy}) is insensitive to a Lorentz rotation of
$E^a(x)$, $E^b(x)$. The connection $\Gamma^a_{\mu b}$ is then completely
analogous to an usual gauge connection, and its Wilson loop
\beq
  {\cal W}(C) = -4 + {\rm Tr} \, ({\cal U}^a_b)(C)
\label{fuz}
\eeq
is a natural invariant quantity of the theory. But from eq.\ (\ref{kxq}) we see
that this loop is equal to that defined in (\ref{fuy}). So the Christoffel
connection $\Gamma^\alpha_{\mu \beta}$ and the anholonomic $\Gamma^a_{\mu b}$
connection have the same loop, denoted by ${\cal W}(C)$. In the computations we
shall employ the connection $\Gamma^\alpha_{\mu \beta}$, which is usually
simpler to deal with.

\m
When the exponential in (\ref{fuy}) is expanded, one obtains terms with 1, 2,
3,
... fields $\Gamma$. We introduce the notation, to be used in the following
\beqa
  {\cal U} & = & {\bf 1} + \oint_C dx^\mu \, \Gamma_\mu(x) +
  \half \, {\rm P} \oint dx^\mu \oint dy^\nu \, \Gamma_\mu(x) \,
  \Gamma_\nu(y) + ... \\
   & = & {\bf 1} + {\cal U}^{(1)} + \frac{1}{2} \, {\cal U}^{(2)} + ...
\eeqa
and
\beq
  {\cal W} = -4 + {\rm Tr} \, {\cal U} =
  {\rm Tr} \, {\cal U}^{(1)} + \half \, {\rm Tr} \, {\cal U}^{(2)} + ...
\eeq


\section{Dynamics.}
\label{dynami}

We shall assume that the dynamics of the gravitational field is given by the
Einstein-Hilbert action
\beq
  S = - \frac{1}{16\pi G} \int d^4x \, \sqrt{g(x)} \, R(x) .
\label{vje}
\eeq
In the vierbein formalism $S$ is expressed as
\beq
  S = - \frac{1}{16\pi G} \int d^4x \,
  R^{ab}_{\mu \nu}(x) e^c_\rho(x) e^d_\sigma(x)
  \epsilon^{\mu \nu \rho \sigma} \epsilon_{abcd} ,
\label{hku}
\eeq
where $R^{ab}_{\mu \nu}$ is the usual gauge curvature of $\Gamma^a_{b \mu}$.

As it is well known, Einstein's gravity written in the form (\ref{hku}) is a
gauge theory of the Lorentz group (i.e., $S$ is invariant under local Lorentz
transformations), but not of the whole Poincar\'e group $ISO(3,1)$. A gauge
formulation can be obtained only introducing some auxiliary
fields $q^a$ \cite{grinar}.

So it is not possible to consider in (3+1) dimensions, like in (2+1)-gravity
\cite{witten}, the holonomies of the Lie algebra valued connection
\beq
  {\cal A}_\mu(x) = e^a_\mu(x) P_a + \Gamma^{ab}_\mu(x) \omega_{ab} ,
\eeq
where $P_a$ and $\omega_{ab}$ are the generators of the translations and of the
Lorentz transformations.

{}From the dynamical point of view, the holonomies of ${\cal A}_\mu$ have
certainly more content than the holonomies of $\Gamma_\mu$ alone. For instance,
it can be easily verified that the term
\beq
  {\rm Tr} \oint dx^\mu \oint dy^\nu \la
  e^a_\mu(x) P_a \, e^b_\nu(y) P_b \ra_0 =
  -2 \delta_{ab} \oint dx^\mu \oint dy^\nu \la
  e^a_\mu(x) \, e^b_\nu(y) \ra_0
\eeq
is not trivial to leading order, unlike the corresponding term containing
the connection (see \S\ \ref{smaqua}). However, this term does not
respect the invariance of the action. In conclusion, the loop ${\cal W}(C)$
defined in eq.s (\ref{fuy}), (\ref{fuz}) is the only invariant loop
functional which we can define in Einstein's gravity.

\m
In the quantum case, we assume the quantum averages to be given by the
functional integral
\beq
  z = \int d[g] \, \exp \frac{i}{\hbar} \left\{ S[g] \right\} ,
\label{dhp}
\eeq
or by the analogous formula in the first order formalism.

A roman letter corresponding to a calligraphic one will
denote the vacuum average of the classical quantity. For instance, we write
\beqa
  U = \la {\cal U} \ra_0 & = & {\bf 1} + \la {\cal U}^{(1)} \ra_0
  + \frac{1}{2} \, \la {\cal U}^{(2)} \ra_0 + ... \nonumber \\
  & = & {\bf 1} + U^{(1)} + \frac{1}{2} \, U^{(2)} + ... \\
  W = \la {\cal W} \ra_0 & = & -4 + {\rm Tr} \, U  \nonumber \\
  & = & {\rm Tr} \, U^{(1)} + \frac{1}{2} {\rm Tr} \, U^{(2)} + ...
  \nonumber \\
  & = & W^{(1)} + \frac{1}{2} W^{(2)} + ...
\eeqa


\section{Classical case.}
\label{clacas}

We just give one typical example of a classical holonomy, namely that
of the Schwarz\-schild solution. Let us consider the Schwarzschild metric
\cite{weinbe}
\beq
  d\tau^2 = B(r) dt^2 - A(r) dr^2 - r^2 \left( d\theta^2
  + \sin^2 \theta \, d\phi^2 \right) ,
\eeq
where
\beq
  B(r) = \left( 1 - \frac{2MG}{r} \right); \qquad \qquad
  A(r) = \left( 1 - \frac{2MG}{r} \right)^{-1} .
\eeq

The corresponding Christoffel connection has the following non-vanishing
components
\beqa
  & & \Gamma^r_{\phi \phi} = - \frac{r \sin^2 \theta}{A(r)}; \qquad \qquad
  \ \ \ \Gamma^\phi_{\phi r} = \frac{1}{r} ; \\
  & & \Gamma^\theta_{\phi \phi} = - \sin \theta \cos \theta ; \qquad \qquad
  \Gamma^\phi_{\phi \theta} = {\rm cotan} \, \theta .
\eeqa

Let us take as the curve $C$ a circle of radius $r_0$, azimut $\theta_0$
and constant time $t_0$; that is, $C$ is described by the function
\beq
  x^\mu(\phi) = \left( t_0,\, r_0,\, \theta_0,\, \phi \right) .
\eeq

The linear term ${\cal W}^{(1)}$ in the holonomy is given by (we omit, for
brevity, the arguments of the field)
\beqa
  {\cal W}^{(1)} & = & \delta^\sigma_\rho \oint dx^\mu \,
  \Gamma^{\rho}_{\mu \sigma} \nonumber \\
  & = & \delta^\sigma_\rho \left\{
  \oint dt \, \Gamma^\rho_{t \sigma} + \oint dr \, \Gamma^\rho_{r \sigma} +
  \oint d\theta \, \Gamma^\rho_{\theta \sigma} +
  \oint d\phi \, \Gamma^\rho_{\phi \sigma} \right\} \nonumber \\
  & = & \oint d\phi \, \left\{ \Gamma^t_{\phi t} + \Gamma^r_{\phi r} +
  \Gamma^\theta_{\phi \theta} + \Gamma^\phi_{\phi \phi} \right\} = 0 .
\eeqa

The quadratic term is
\beqa
  {\cal W}^{(2)} & = & {\rm P} \, \delta^\gamma_\alpha
  \oint dx^\mu \oint dy^\nu \, \Gamma^\alpha_{\mu \beta}
  \, \Gamma^\beta_{\nu \gamma} \nonumber \\
  & = & {\rm P} \oint d\phi' \oint d\phi'' \,
  \Gamma^\alpha_{\phi' \beta} \, \Gamma^\beta_{\phi'' \alpha} \nonumber \\
  & = & 2 \int_0^{2\pi} d\phi' \int_0^{\phi'} d\phi'' \, \left\{
  \Gamma^r_{\phi' \phi} \, \Gamma^\phi_{\phi'' r} +
  \Gamma^\theta_{\phi' \phi} \, \Gamma^\phi_{\phi'' \theta}
  + \Gamma^\phi_{\phi' r} \, \Gamma^r_{\phi'' \phi} +
  \Gamma^\phi_{\phi' \theta} \, \Gamma^\theta_{\phi'' \phi}
  \right\} \nonumber \\
  & = & - 8 \pi^2 \left( \frac{\sin^2 \theta_0}{A(r_0)} + \cos^2 \theta_0
  \right)
\label{guy}
\eeqa

It is easy to check that the contribution ${\cal W}^{(3)}$, of the form
\beq
  \oint d\phi' \oint d\phi'' \oint d\phi'''
  \Gamma^\alpha_{\phi' \beta} \Gamma^\beta_{\phi'' \gamma}
  \Gamma^\gamma_{\phi''' \alpha} ,
\eeq
vanishes, and the same do the following contributions. Thus the total
holonomy is exactly given by
\beq
  {\cal W} = - (2\pi)^2 \left( \frac{\sin^2 \theta_0}{A(r_0)} + \cos^2 \theta_0
  \right) \simeq - (2\pi)^2 \left( 1 + \frac{2MG \sin^2 \theta_0}{r_0} \right)
\eeq

We see that if we set $\theta_0=0$ (``equatorial'' circle) the loop is constant
and equal to $- (2\pi)^2$ (for symmetry reasons); if we set
$\theta_0 \neq 0$, we have a small ``precession angle'' (see \S\ \ref{geoint})
which depends on $r_0$ and vanishes when $r_0 \to \infty$.


\section{Small quantum fluctuations around a flat background.}
\label{smaqua}

In this case, following the usual approach, we decompose the metric $g_{\mu
\nu}(x)$ as
\beq
  g_{\mu \nu}(x) = \eta_{\mu \nu} + \kappa h_{\mu \nu}(x);
  \qquad \kappa = \sqrt{16\pi G} ,
\label{kuy}
\eeq
and we interpret $\eta_{\mu \nu}$ as the classical background, while $\kappa
h_{\mu \nu}(x)$ is regarded as a small quantized perturbation, which represents
gravitons propagating in the vacuum. The Einstein-Hilbert lagrangian
(\ref{vje}) is then splitted into a quadratic part, whose inverse gives the
bare graviton propagator, and into interaction vertices. Due to the
non polynomial character of the lagrangian, there are infinitely many vertices;
the first two ones, respectively proportional to $\kappa$ and $\kappa^2$,
connect 3 and 4 fields $h$. Hence the first few orders of perturbation theory
are formally very similar to those of Yang-Mills theory.

\m
The leading contribution to $W$, of order $\hbar \kappa^2$, is given by
$W^{(2)}$ with one bare propagator, namely
\beq
  W^{(2)} = \oint_C dx^\mu \oint_C dy^\nu
  \la \Gamma^\beta_{\mu \alpha}(x) \, \Gamma^\alpha_{\nu \beta}(y) \ra .
\label{dle}
\eeq
Here the brackets denote the bare propagator of the $\Gamma$'s, obtained using
their definition (\ref{ces}), eq.\ (\ref{kuy}) and the propagator of $h_{\mu
\nu}(x)$ (see \S\ \ref{stagra}, eq.\ (\ref{ppi})).

The following two contributions to $W$, of order $\hbar^2 \kappa^4$, are given
by the term $W^{(4)}$ with two bare propagators and by the term $W^{(3)}$ with
three propagators and one $\kappa$-vertex. Finally, the three contributions of
order $\hbar^3 \kappa^6$ are given by the term $W^{(6)}$ with three
propagators, by the term $W^{(4)}$ with four propagators and one
$\kappa^2$-vertex and by the term $W^{(2)}$ with the radiatively corrected
propagator.

\m
What is remarkable, and easily shown \cite{gtrips}, is that the leading term
(\ref{dle}), of order $\hbar$, vanishes in Einstein's theory. (This opened the
problem of finding a gauge invariant expression for the static gravitational
potential; see \cite{energy}.) In the following of this Section, we would like
to show that this vanishing, in fact, does not depend on the dynamic content of
Einstein's action, but is only due to the symmetries of the propagator, to
the Poincar\'e invariance of the background and to the absence of a dimensional
coupling (apart from the overall factor $\kappa^{-2}$) in the action
(\ref{vje}).

Let us write the most general form of the propagator $\la h_{\mu \nu}(x)
h_{\alpha \beta}(y) \ra$ which is compatible with the symmetries in the indices
and with Poincar\'e invariance (in any dimension $N$). We have
\beqa
  \la h_{\mu \nu}(x) h_{\alpha \beta}(y) \ra & = &
  a \, \frac{\Delta_{\mu \nu \alpha \beta}}{X^{(N-2)}}
  + b \, \frac{\eta_{\mu \nu} \eta_{\alpha \beta}}{X^{(N-2)}} + \nonumber \\
  & & + c \, \frac{s_{\mu \nu \alpha \beta}}{X^N}
  + d \, \frac{S_{\mu \nu \alpha \beta}}{X^N}
  + e \, \frac{X_\mu X_\nu X_\alpha X_\beta}{X^{(N+2)}} ,
\label{wte}
\eeqa
where
\beqa
  X_\mu & = & x_\mu-y_\mu ; \qquad
  X^N=[(\vec{x}-\vec{y})^2-(x^0-y^0)^2-i\epsilon]^{\frac{N}{2}}; \label{aaa} \\
  \Delta_{\mu \nu \alpha \beta} & = & \half \left(
  \eta_{\mu \alpha} \eta_{\nu \beta} + \eta_{\mu \beta}
  \eta_{\nu \alpha} \right) ; \\
  s_{\mu \nu \alpha \beta} & = & \left(
  \eta_{\mu \nu} X_\alpha X_\beta + \eta_{\alpha \beta} X_\mu X_\nu
  \right) ; \\
  S_{\mu \nu \alpha \beta} & = & \left(
  \eta_{\mu \alpha} X_\nu X_\beta + \eta_{\mu \beta} X_\nu X_\alpha
  + \eta_{\nu \alpha} X_\mu X_\beta + \eta_{\nu \beta}
  X_\mu X_\alpha \right) . \label{bbb}
\eeqa

The tensor $\Delta_{\mu \nu \alpha \beta}$ is the generalization of
$\eta_{\mu \nu}$ to tensors with a symmetric couple of symmetric
indices, and also the tensors $s_{\mu \nu \alpha \beta}$ and $S_{\mu \nu \alpha
\beta}$ are defined in such a way that the decomposition (\ref{wte}) is left
invariant by the exchange of the pair $(\alpha \beta)$ with $(\mu \nu)$ and of
the indices inside each pair.

In (\ref{wte}), $a,\, b,\, c,\, d,\, e\, $ are numerical constants. No other
terms can be present, since there are no other dimensional parameters in the
linearized action. The contribution of order $\hbar \kappa^2$ to the holonomy
is
\beqa
  W^{(2)} & = & \oint dx^\mu \oint dy^\nu \la
  \Gamma^\alpha_{\mu \beta}(x) \Gamma^\beta_{\nu \alpha}(y) \ra \nonumber \\
  & = & \frac{1}{4} \oint dx^\mu \oint dy^\nu \left<
  \left\{ \pa_\beta h^\alpha_\mu(x) - \pa^\alpha h_{\mu \beta}(x) \right\}
  \left\{ \pa_\alpha h^\beta_\nu(y) - \pa^\beta h_{\nu \alpha}(y) \right\}
  \right> \\
  & = & \half \oint dx^\mu \oint dy^\nu \left\{
  \eta^{\alpha \beta} \Box \la h_{\mu \alpha}(x) h_{\nu \beta}(y) \ra -
  \pa^\alpha \pa^\beta \la h_{\mu \alpha}(x) h_{\nu \beta}(y) \ra \right\} .
\label{bka}
\eeqa
It is straightforward to verify that the substitution of (\ref{wte}) into
(\ref{bka}) gives rise only to terms which either are gradients, or ultra-local
terms (that is, containing $\delta^4(x-y)$), or finally are proportional to the
following functions
\beq
  \eta_{\mu \nu} \, \pa^\alpha \pa^\beta
  \, \frac{X_\alpha X_\beta}{X^N}, \qquad \pa^\alpha \pa^\beta \,
  \frac{X_\mu X_\nu X_\alpha X_\beta}{X^{(N+2)}},
\eeq
\beq
  \pa^\alpha \, \frac{X_\mu X_\nu X_\alpha}{X^{(N+2)}} ,
\eeq
which vanish by homogeneity.

\m
As it was pointed out in \cite{correl}, if we admit dimensional couplings in
the action, like in $(R+R^2)$-gravity, some non-vanishing contribution to
$W^{(2)}$ may arise.

Finally, we would like to justify our omission of higher order calculations by
observing that the vanishing of the leading term has a geometrical
interpretation which strongly affects the physical significance of the
holonomies (see \S\ \ref{geoint}). Furthermore, higher order calculations in
quantum gravity are very complicated, and give rise to non renormalizable
infinities, which require the introduction of some effective cut-off. What
would thus seem more appropriate to us, and is in progress now, is to apply
higher order perturbation technique to the formula for the static potential
\cite{energy}.

In the next Section, instead, we shall give a kind of semiclassical expression
for the Wilson loops.


\section{Non-flat background.}
\label{nonfla}

The discussion of the preceding Section suggests that a contribution to the
holonomy proportional to $\hbar$ could arise on a non-flat background. In order
to illustrate this point, let us suppose that a weak external source $J$ for
the gravitational field is present. The field produced by this source, as given
by the Einstein equations, will be denoted, in the variable $h$ defined in
(\ref{kuy}), by $h_{0,\mu \nu}(x)$. The functional integral (\ref{dhp}) will
now depend on $J$: omitting the indices of the field, it is given by
\beq
  z[J] = \int d[h] \, \exp \frac{i}{\hbar} \left\{ S[h] +
  \int dx \, h(x) J(x) \right\} .
\eeq

If we expand the action $S[h]$ around the classical solution $h_0$, we find
\beqa
  S[h] & = & S[h_0] + \int dx \, \left[ \frac{\delta S}{\delta h(x)}
  \right]_{h=h_0} \hat{h}(x) + \nonumber \\
  & & + \, \half \int dx \int dy \,
  \left[ \frac{\delta^2 S}{\delta h(x) \delta h(y)}
  \right]_{h=h_0} \hat{h}(x) \hat{h}(y) + S_3 ,
\eeqa
where
\beq
  h = h_0 + \hat{h} .
\eeq

Since by definition we have
\beq
  \left[ \frac{\delta S}{\delta h(x)} \right]_{h=h_0} = - J(x) ,
\eeq
we are left with \cite{jackiw}
\beqa
  z[J] & = & \exp \frac{i}{\hbar} \left\{ S[h_0] +
  \int dx \, h_0(x) J(x) \right\} \times \nonumber \\
  & & \times \int d[\hat{h}] \exp \frac{i}{2\hbar} \left\{
  \int dx \int dy \left[ \frac{\delta^2 S}{\delta h(x) \delta h(y)}
  \right]_{h=h_0} \, \hat{h}(x) \hat{h}(y) + S_3 \right\} .
\eeqa

The operator
\beq
  G(x,y) = \left\{ \left[ \frac{\delta^2 S}{\delta h(x) \delta h(y)}
  \right]_{h=h_0} \right\}^{-1}
\eeq
is the graviton propagator in the background $h_0$. If we write
symbolically the Einstein action as a quadratic part $Q$ plus the
interaction vertices $V^{(3)}$ and $V^{(4)}$, in the following way
\beq
  S[h] = \half Q h^2 + \frac{1}{6} \kappa V^{(3)} h^3 +
  \frac{1}{12} \kappa^2 V^{(4)} h^4 + O(\kappa^3) ,
\eeq
we have
\beq
  \left[ \frac{\delta^2 S}{\delta h^2} \right]_{h=h_0} =
  Q + \kappa V^{(3)} h_0 + \kappa^2 V^{(4)} h_0^2 + O(\kappa^3) ,
\eeq
whose inverse is
\beq
  G = Q^{-1} - \kappa V^{(3)} h_0 + O(\kappa^2) ,
\label{cfs}
\eeq
where $Q^{-1}$ is the usual propagator of the graviton on a flat background.
When evaluating the holonomy, we have to compute the expectation value
\beq
  W = \frac{\int d[\hat{h}] \, \exp \frac{i}{2\hbar} \left\{ \int dx \int dy
  \, G^{-1}(x,y) \hat{h}(x) \hat{h}(y) + S_3 \right\} {\cal W}[h_0+\hat{h}]}
  {\int d[\hat{h}] \, \exp \frac{i}{2\hbar} \left\{ \int dx \int dy
  \, G^{-1}(x,y) \hat{h}(x) \hat{h}(y) + S_3 \right\}} .
\eeq

It is known that $S_3$ is of higher order in $\hbar$; thus the contribution of
order $\hbar$ to $W$ is still given by eq.\ (\ref{dle}), where the propagator
is now given by (\ref{cfs}). The term with $Q^{-1}$ vanishes, as we saw in the
preceding Section; the other term in general does not vanish, and gives a
contribution to the holonomy proportional to $\hbar \kappa^3$.

Thus we have seen that breaking the Poincar\'e invariance with a small source
term which produces a non-flat background, we may obtain a contribution
to the quantum holonomies proportional to $\hbar$, while there is no such
contribution on a flat background. Nevertheless, this is a small effect,
being proportional to $\kappa^3$.


\section{Geometrical and physical interpretation.}
\label{geoint}

It is interesting at this point to do a sharper analysis of the properties of
the matrix ${\cal U}(C)$ of the parallel transport defined in \S\ \ref{defini}.
We shall see that in the euclidean theory the vanishing of its trace amounts to
a very strong geometrical statement.

Let us first consider, for illustrative purposes, the case of a Yang-Mills
theory of the group $SO(3)$. The gauge connection has the form
\beq
  A_\mu(x) = A^i_\mu(x) \, L_i \, ; \qquad i=1,2,3 ,
\eeq
where the matrices $L_i$ constitute a representation of the Lie algebra of the
group. In particular, to fix the ideas, let us choose the adjoint
representation; in this case the matrices $L_i$ have elements $(L_i)^m_n$
($m,n=1,2,3$), which are related to the structure constants $\varepsilon_{imn}$
of the group. The connection $A_\mu(x)$ performs the parallel transport of a
3-dimensional vector $V^n$ in the ``internal'' space according to the formula
(compare eq.\ (\ref{xnc}))
\beq
  dV^m = A^i_\mu(x) \, (L_i)^m_n \, V^n \, dx^\mu .
\eeq
The vector is rotated during the transport, but its length remains unchanged.
Let us consider the matrix ${\cal O}(C)$ which describes the parallel transport
along a closed curve $C$. ${\cal O}(C)$ is defined by a P-exponential, through
a formula similar to eq.\ (\ref{fuy}). Suppose that we take a vector $V$ in a
point $P$ of $C$, and parallel-transport it along $C$, returning to $P$; let us
denote by $V'$ the new vector we obtain in this way. The vectors $V$ and $V'$
have the same length, that is
\beq
  \delta_{mn}V^m V^n=\delta_{mn}{V'\, }^m {V'\, }^n ,
\label{cbr}
\eeq
but they differ by an angle $\theta$, which is related to the trace of ${\cal
O}(C)$. For small angles, we have, by a proper choice of the coordinate axes in
the internal space
\beq
  {\cal O}(C) = \left(
  \begin{array}{ccc}
    1 - \half \theta^2 & \theta & 0 \\
    - \theta & 1 - \half \theta^2 & 0 \\
    0 & 0 & 1
  \end{array}
  \right) ,
\label{cty}
\eeq
that means
\beq
  {\rm Tr} \, {\cal O}(C) = 3 - \theta^2 .
\eeq

More generally, we remind that the Lie algebra of $SO(3)$ has just one Casimir
invariant, namely the operator
\beq
  L^2 = L_1^2 + L_2^2 + L_3^2 .
\eeq
This operator commutes with each of the $L_i$'s, so we can in general rotate
our coordinate system as to have $L^2 = L_3^2$, and the rotation matrix takes
in this case the form (\ref{cty}), i.e.\ we have
\beq
  {\cal O}(C)={\bf 1}+\theta L_3 + \half \theta^2 L_3^2 + ...
\label{cvp}
\eeq
Taking the trace of (\ref{cvp}), remembering that ${\rm Tr}\, L_i=0$
and using the normalization condition of the Lie generators
\beq
  {\rm Tr} \, L_i L_j = - 2 \delta_{ij} ,
\eeq
we find that $\theta^2$ is the coefficient of the Casimir invariant in the
expansion of the exponential.

\m
Next we come to consider the group $SO(4)$. Intuitively, adding a new dimension
we can make an independent rotation. Multiplying two 4-dimensional matrices
similar to (\ref{cty}), the first representing a rotation by an angle
$\theta_I$ perpendicular to one plane and the second a rotation by another
angle $\theta_{II}$ perpendicular to another plane, we find that
\beq
  {\rm Tr} \, {\cal O}(C) = 4 - ( \theta_I^2 + \theta_{II}^2 )
\label{vde}
\eeq
Also we know that $SO(4)=[SO(3)]_I \, \times \, [SO(3)]_{II}$ and that
we have two Casimirs now \cite{wybour}, corresponding to $(L_I^2 + L_{II}^2)$,
whose ``eigenvalue'' appears in (\ref{vde}), and $(L_I^2 - L_{II}^2)$,
which is not of interest in this case.

\m
The group $SO(4)$ is the relevant one for euclidean quantum gravity. In fact,
the geometrical interpretation of the matrix ${\cal U}(C)$ is the following.
During the parallel transport of a vector $V$ in spacetime, its length, given
by
\beq
  |V|^2 = V^a V^b \delta_{ab} = V^\mu V^\nu g_{\mu \nu}(x),
\eeq
does not change. If we transport $V$ along a closed curve $C$, returning
to the starting point, we obtain another vector $V'$, which has the same
length of $V$ , and differs from it only in the orientation. Hence we have
for any vector
\beq
  V^a V^b \delta_{ab} = {V'\,}^a {V'\,}^b \delta_{ab} =
  {\cal U}^a_c(C) V^c \, {\cal U}^b_d(C) V^d \, \delta_{ab} ,
\eeq
or, in matrix notation,
\beq
  {\cal U}^T(C) \, {\cal U}(C) = {\bf 1} .
\eeq
The matrix ${\cal U}$ belong then to $SO(4)$ and its trace has the form
(\ref{vde}).

\m
If the variance of the angles $\theta_I$ and $\theta_{II}$ is zero
to order $\hbar$ (because $W^{(2)}$ vanishes), the angles
themselves have to vanish identically in any configuration, that is
\beq
  {\cal U}(C) = {\bf 1} \ \ {\rm for \ any} \ C.
\eeq
This is a very strong geometrical statement, as it implies that, still
to order $\hbar$, all the weak field configurations which effectively enter
the functional integral
\beq
  z = \int d[h] \, \exp \left\{ - \hbar^{-1} \, S[h] \right\}
\label{dks}
\eeq
have no curvature. In other words, the curved configurations -- which possibly
dominate in other regimes -- are in this approximation totally suppressed.

\m
This unexpected situation should be compared with what happens, for instance,
in a ordinary $SO(3)$ or $SO(4)$ gauge theory. In this case the leading term
$W^{(2)}(C)$ does not vanish and the variance of the rotation angles is not
zero to order $\hbar$.  For instance, if the curve $C$ has the form of a
rectangle of sides $L$ and $T$, with $L \ll T$, the quantity $\, -(\hbar
T)^{-1} \log \la \theta^2 \ra_0$ is the potential energy of two non-abelian
charges kept at rest at a distance $L$ each from the other.

So the matrices of the parallel transport in the ``internal'' gauge manifold,
considered configuration by configuration, are not equal to the identity
matrix. Interpreting $\hbar$ as the temperature $\Theta$ of an equivalent
statistical system, we see that when $\Theta$ grows from zero to some small
value -- such that we may disregard $\Theta^2$ or higher orders -- the
Yang-Mills fields develop ``localized excitations'', i.e.\ regions of various
sizes where the Yang-Mills curvature is not vanishing.

All this does not happen for the gravitational field, which remains essentially
in a ``flat'' state. Such a picture also explains the absence in this
approximation of any invariant correlation of the curvature \cite{correl}.

\m
We also have seen that the introduction of a small external
source in the functional integral (\ref{dks}), breaking the Poincar\'e
invariance of the background, gives rise in general to a non vanishing
contribution to the loop proportional to $\hbar$. In this case we may have
excitations with localized curvature, but they are very small, since their
variance is proportional to $\kappa^3$ instead of $\kappa^2$ (they are in fact
originated by the non-linear interaction of gravitons).


\section{Stabilized gravity.}
\label{stagra}

In a series of papers \cite{greens}, Greensite has recently proposed a new
``stabilized'' action for euclidean quantum gravity. It is known \cite{hawkin}
that the euclidean action obtained from Einstein's action by {\it naive}
analytical continuation is not bounded from below, due to the ``wrong sign'' of
the conformal term in the kinetic operator. On the other hand, it is not
obvious in quantum gravity that the simple analytical continuation is a correct
procedure.

Both field-theoretical work \cite{mazurm} and a suitably modified stochastic
quantization procedure \cite{greens} suggest that in the ``right'' euclidean
action the sign of the conformal factor is flipped to lowest order, while to
higher orders the action itself becomes non-local.

In the following of this Section, also in view of future applications,
we shall find the propagator of stabilized gravity and verify that it
gives the expected result for the holonomies to leading order.

According to the notation of ref.\ \cite{greens}, the linearized euclidean
gravitational action is written as
\beq
  S^0 = \int \frac{d^4p}{(2\pi)^4} \, \tilde{h}_{\mu \nu}(p) \, p^2
  \, \tilde{K}_{\mu \nu \alpha \beta}(p) \,
  \tilde{h}_{\alpha \beta}(p) .
\eeq
In the usual Einstein theory the kinetic operator $\tilde{K}$ is given by
\beq
  \tilde{K}^{\rm Einstein}_{\mu \nu \alpha \beta} =
  P^{(2)}_{\mu \nu \alpha \beta} -
  2 P^{(0-s)}_{\mu \nu \alpha \beta} ,
\label{egr}
\eeq
where $P^{(2)}$ and $P^{(0-s)}$ are the projection operators
\beqa
  P^{(2)}_{\mu \nu \alpha \beta} & = &
  \half \left( \theta_{\mu \alpha} \theta_{\nu \beta} +
  \theta_{\mu \beta} \theta_{\nu \alpha} \right)
  - \frac{1}{3} \theta_{\mu \nu} \theta_{\alpha \beta} ; \\
  P^{(0-s)}_{\mu \nu \alpha \beta} & = &
  \frac{1}{3} \theta_{\mu \nu} \theta_{\alpha \beta} ; \\
  \theta_{\mu \nu} & = & \delta_{\mu \nu} - \frac{1}{p^2} p_\mu p_\nu .
\eeqa

The kinetic operator of the linearized effective action of stabilized gravity
is simply obtained by changing the sign in (\ref{egr})
\beq
  \tilde{K}^{\rm Stabilized}_{\mu \nu \alpha \beta} =
  P^{(2)}_{\mu \nu \alpha \beta} +
  2 P^{(0-s)}_{\mu \nu \alpha \beta} .
\label{sfr}
\eeq

Explicit evaluation of $\tilde{K}^{\rm Einstein}$ leads to the expression
\beq
  \tilde{K}^{\rm Einstein}_{\mu \nu \alpha \beta} =
  \Delta_{\mu \nu \alpha \beta} - \delta_{\mu \nu}
  \delta_{\alpha \beta} + \frac{1}{p^2} \tilde{s}_{\mu \nu \alpha \beta}
  - \frac{1}{2p^2} \tilde{S}_{\mu \nu \alpha \beta} ,
\label{jdf}
\eeq
where the tensors $\Delta_{\mu \nu \alpha \beta}$,
$\tilde{s}_{\mu \nu \alpha \beta}$ and $\tilde{S}_{\mu \nu \alpha \beta}$
are the analogues in $p$-space of those defined in eq.s (\ref{aaa}) --
(\ref{bbb}).

For $\tilde{K}^{\rm Stabilized}$ we have instead, expanding (\ref{sfr}) ,
\beq
  \tilde{K}^{\rm Stabilized}_{\mu \nu \alpha \beta} =
  \Delta_{\mu \nu \alpha \beta} + \frac{1}{3}
  \, \delta_{\mu \nu} \delta_{\alpha \beta} -
  \frac{1}{3p^2} \, \tilde{s}_{\mu \nu \alpha \beta} -
  \frac{1}{2p^2} \, \tilde{S}_{\mu \nu \alpha \beta}
  + \frac{4}{3p^4} \, p_\mu p_\nu p_\alpha p_\beta .
\eeq

\m
The kinetic operators above are not invertible. In order to find the
corresponding propagators, we must add to them a gauge-fixing term, usually
the harmonic gauge fixing
\beq
  \tilde{K}^{\rm Harmonic}_{\mu \nu \alpha \beta} =
  \frac{1}{2} \delta_{\mu \nu} \delta_{\alpha \beta} -
  \frac{1}{p^2} \tilde{s}_{\mu \nu \alpha \beta} + \frac{1}{2p^2}
  \tilde{S}_{\mu \nu \alpha \beta} .
\eeq

Then we consider the following propagator equation
\beq
  p^2 \left[ \tilde{K}_{\mu \nu \alpha \beta}(p)
  + \tilde{K}^{\rm Harmonic}_{\mu \nu \alpha \beta}(p) \right]
  \, \tilde{G}_{\alpha \beta \rho \sigma}(p) =
  - \Delta_{\mu \nu \rho \sigma}
\eeq
and look for a solution of the general form
\beq
  \tilde{G}_{\alpha \beta \rho \sigma} =
  \frac{a}{p^2} \, \Delta_{\alpha \beta \rho \sigma}
  + \frac{b}{p^2} \, \delta_{\alpha \beta} \delta_{\rho \sigma}
  + \frac{c}{p^4} \, \tilde{s}_{\alpha \beta \rho \sigma}
  + \frac{d}{p^4} \tilde{S}_{\alpha \beta \rho \sigma}
  + \frac{e}{p^6} \, p_\alpha p_\beta p_\rho p_\sigma ,
\eeq
where $a,\, b,\, c,\, d,\, e\, $ are numerical constants.

In the case of Einstein's theory we find
\beq
  \left( a^{\rm Ei.}=-1,\ b^{\rm Ei.}=\half ,\ c^{\rm Ei.}=0,
  \ d^{\rm Ei.}=0,\ e^{\rm Ei.}=0 \right) .
\eeq
which corresponds in the $x$-space to the familiar Feynman-De Witt propagator
\cite{veltma}
\beq
  \la h_{\mu \nu}(x) h_{\rho \sigma}(y) \ra^{\rm Ei.} =
  - \frac{\hbar}{8\pi^2}
  \frac{\delta_{\mu \rho} \delta_{\nu \sigma} + \delta_{\mu \sigma}
  \delta_{\nu \rho} - \delta_{\mu \nu} \delta_{\rho \sigma}}{(x-y)^2} .
\label{ppi}
\eeq

In the case of stabilized gravity the solution is
\beq
  \left(a^{\rm St.}=-1,\ b^{\rm St.}=\frac{1}{6},
  \ c^{\rm St.}=-\frac{2}{3},
  \ d^{\rm St.}=0,\ e^{\rm St.}=-\frac{4}{3} \right) .
\eeq

In order to write the corresponding $x$-space propagator, we must compute
the Fourier transforms of the non-standard terms of the form $p^{-4}p_\alpha
p_\beta$ and $p^{-6}p_\alpha p_\beta p_\rho p_\sigma$. This computation is done
in details in the Appendix. The result is pretty simple:
\beqa
  \int \frac{d^4p}{(2\pi)^4} \, \frac{p_\alpha p_\beta}{p^4} \, e^{ipx} & = &
  \frac{1}{(2\pi)^2} \, \frac{x_\alpha x_\beta}{x^4} ;
  \label{cka} \\
  \int \frac{d^4p}{(2\pi)^4} \,
  \frac{p_\alpha p_\beta p_\rho p_\sigma}{p^6} \, e^{ipx} & = &
  \frac{1}{(2\pi)^2} \, \frac{x_\alpha x_\beta x_\rho x_\sigma}{x^6} .
  \label{bht}
\eeqa

So the propagator is (with $X=x-y$)
\beqa
  \la h_{\mu \nu}(x) h_{\rho \sigma}(y) \ra^{\rm St.} & = &
  - \frac{1}{(2\pi)^2 X^2} \, \Delta_{\mu \nu \rho \sigma} +
  \frac{1}{6(2\pi)^2 X^2} \, \delta_{\mu \nu} \delta_{\rho \sigma}
  \nonumber \\
  & & - \, \frac{2}{3(2\pi)^2 X^4} \, s_{\mu \nu \rho \sigma} -
  \frac{4}{3(2\pi)^2 X^6} \, X_\mu X_\nu X_\rho X_\sigma .
\eeqa

Being of the form (\ref{wte}) it gives no contribution to the leading
term of the holonomies (see \S\ \ref{smaqua}).

\section{Concluding remarks.}
\label{conclu}

In this work the behavior of quantum and semiclassical Wilson loops has been
studied perturbatively in four-dimensional pure Einstein gravity. The main
results comprise the vanishing of the leading perturbative contribution to the
loops and a geometrical interpretation of this vanishing in terms of the
structure of the vacuum state. We also have treated the case of a non flat
background and that of stabilized gravity.

The most interesting ``discovery'' of our analysis, from the physical point of
view, is that the vacuum state of quantum gravity does not show, to order
$\hbar$, any field configuration with localized curvature. This behavior is
very different from that of other gauge fields.

\m
But if the Wilson loops vanish and if there is no localized curvature, how can
we express in a invariant way the interaction energy of two masses, and how can
we think of the ``mechanism'' of their gravitational interaction?

The first question has a definite formal answer, in terms of an essentially
non-local formula \cite{energy}.

The second question is more subtle -- also in view of the difficulties
encountered already at the classical level for the definition of a localized
gravitational energy (see \cite{energy}). All we can say at the present stage
is that the mechanism seems to be not strictly local. It could be possible to
find some analogue of this behavior in other field models; work is in progress
in this direction.

\m
One limit of our analysis resides in its perturbative nature. Of course, non
perturbative analyses of quantum gravity are a major challenge. Nevertheless,
all the considerations above do not involve particularly short distances, where
gravity is thought to develope highly non-perturbative features. As we pointed
out in the Introduction, perturbative quantum gravity may be regarded in our
case just as an effective field theory.

\section{Acknowledgements.}

I would like to thank Prof.\ Roman Jackiw for his kind hospitality at M.I.T.\
and for very invaluable discussions. I also have benefited from helpful
discussions with D.\ Cangemi, D.Z.\ Freedman and C.\ Lucchesi and from an
earlier conversation with M.\ Toller.

The author is supported by a fellowship of the Foundation ``A.\ Della Riccia''
of Florence, Italy.

\vskip 2 cm

\centerline{\bf APPENDIX.}

\bigskip
We want to prove eq.s (\ref{cka}), (\ref{bht}), namely
\beqa
  \int \frac{d^4p}{(2\pi)^2} \, \frac{p_\alpha p_\beta}{p^4} \, e^{ipx} & = &
  \frac{x_\alpha x_\beta}{x^4} ;
  \label{ckk} \\
  \int \frac{d^4p}{(2\pi)^2} \,
  \frac{p_\alpha p_\beta p_\rho p_\sigma}{p^6} \, e^{ipx} & = &
  \frac{x_\alpha x_\beta x_\rho x_\sigma}{x^6} ,
  \label{bhh}
\eeqa
starting from the known result
\beq
  \int \frac{d^4p}{(2\pi)^2} \, \frac{e^{ipx}}{p^2} =
  \frac{1}{x^2} .
\eeq

Let us first consider (\ref{ckk}). By euclidean invariance we will have
\beq
  \int \frac{d^4p}{(2\pi)^2} \, \frac{p_\alpha p_\beta}{p^4} \, e^{ipx} =
  A \, \frac{\delta_{\alpha \beta}}{x^2} +
  B \, \frac{x_\alpha x_\beta}{x^4} ,
\eeq
where $A$ and $B$ are two unknown coefficients, which we may determine
imposing the following conditions:
\beqa
  & & \delta^{\alpha \beta}
  \left( A \, \frac{\delta_{\alpha \beta}}{x^2} +
  B \, \frac{x_\alpha x_\beta}{x^4} \right)=
  \frac{1}{x^2} ; \label{nye} \\
  & & \int \frac{d^4x}{(2\pi)^2} \, \left( A \,
  \frac{\delta_{\alpha \beta}}{x^2} +
  B \, \frac{x_\alpha x_\beta}{x^4} \right) \, e^{-ipx}
  = \frac{p_\alpha p_\beta}{p^4} . \label{dkw}
\eeqa
Eq.\ (\ref{dkw}) express the invertibility of the Fourier transform.
We obtain two possible solutions:
\beq
  \left( A=\half , \ B=-1 \right) \qquad
  {\rm and} \qquad \left( A=0 , \ B=1 \right) .
\eeq

Then we write
\beqa
  \int \frac{d^4p}{(2\pi)^2} \, \frac{p_\alpha p_\beta p_\rho p_\sigma}{p^6}
  \, e^{ipx} & = & \frac{a}{x^2} \left( \delta_{\alpha \beta}
  \delta_{\rho \sigma} + ... \right) + \nonumber \\
  & & + \frac{b}{x^4} \left( \delta_{\alpha \beta} x_\rho x_\sigma
  + ... \right) + \frac{c}{x^6} \, x_\alpha x_\beta x_\rho x_\sigma ,
\eeqa
where the dots denote all the possible symmetrizations, and impose the
conditions analogous to (\ref{nye}), (\ref{dkw}). In this way we find that a
solution for $(a,b,c)$ exists only if $(A=0, \, B=1)$ and in this case we have
\beq
  \left( a=0, \ b=0, \ c=1 \right) .
\eeq


\end{document}